\documentclass[submitting]{nst}

\usepackage{subfigure,dcolumn}
\usepackage{epstopdf}
\usepackage{mhchem}
\usepackage{upgreek}
\usepackage{framed}
\usepackage{booktabs}
\usepackage{listings}
\usepackage{hyperref}

\let\linenumbers\relax

\begin{document}

\title{Experimental Dataset from BESIII detector at Beijing Electron-Positron Collider} 
\thanks{This work was supported by the National Key Research and Development Program of China (Grant No.~2023YFA1606000), National Natural Science Foundation of China (Grant Nos.~12175321, 11975021, and U1932101), and National College Students Science and Technology Innovation Project of Sun Yat-sen University.}

\author{Ming-Hua Liao}
\affiliation{School of Physics, Sun Yat-Sen University, Guangzhou 510275, China}
\author{Jian-Shu Liu}
\affiliation{School of Physics, Sun Yat-Sen University, Guangzhou 510275, China}
\author{Xin-Nan Wang}
\affiliation{Institute of High Energy Physics, Chinese Academy of Sciences, Beijing 100049, China}
\author{Sheng-Sen Sun}
\affiliation{Institute of High Energy Physics, Chinese Academy of Sciences, Beijing 100049, China}
\affiliation{University of Chinese Academy of Sciences, Beijing, 100049, China}
\author{Zheng-Yun You}
\email[Zheng-Yun You, ]{youzhy5@mail.sysu.edu.cn}
\affiliation{School of Physics, Sun Yat-Sen University, Guangzhou 510275, China}

\begin{abstract}
In the BESIII detector at Beijing electron-positron collider, billions of events from $e^{+}e^{-}$ collisions are recorded. These events passing through the trigger system are saved in raw data format files. They play an important role in the study of physics at~$\tau$-charm energy region. Here, we publish an $e^{+}e^{-}$ collision dataset containing both Monte Carlo simulation samples and real data collected by the BESIII detector. The data passes through the detector trigger system, file format conversion, physics information extraction, and finally saves the physics information and detector response in text format files. This dataset is publicly available and is intended to provide interested scientists and those outside of the BESIII collaboration with event information from BESIII that can be used in understanding the physics research in $e^{+}e^{-}$ collisions, developing visualization projects for physics education, public outreach, and science advocacy.

\end{abstract}


\maketitle

\begin{table*}[!htb]
Specifications Table \\
\begin{tabular*}{\hsize} {@{\extracolsep{\fill} } ll}
\toprule
Subject & Particles Physics \\ 
Specific subject area & Electron-Positron collider experiment.\\
Data format & Raw\&Analyzed\\
Type of data & Text format file\\
How data were acquired & Measurements by the BESIII detector and processed with BESIII offline software. \\
Parameters for data collection & Electron-positron collisions at different energy points. \\
Description of data collection& Data were collected by the BESIII detector and reconstructed for physics analysis. \\
Data collection & The data were collected from electron-positron collisions at BEPCII using the BESIII detector. \\
Data source location & Institution: Institute of High Energy Physics, Chinese Academy of Sciences\\
                                 &Country: China\\
Data accessibility& Repository name: Science Data Bank\\
                            &Data identification number: https://cstr.cn/31253.11.sciencedb.21486\\
                            &Direct URL to data: https://doi.org/10.57760/sciencedb.21486 \\
Related research article & Z.~J.~Li, et al. Front. Phys. (Beijing) \textbf{19} (2024) no.6, 64201 doi:10.1007/s11467-024-1422-7 \\
\bottomrule
\end{tabular*}
\end{table*}

\section{Introduction}\label{sec.I}

Beijing Electron-Positron Collider~(BEPC)~\cite{Ye:1987nh} has operated since 1989 and completed the upgrade to BEPCII in 2008. BEPCII was designed to operate at $\tau$-charm energy region. It is a double-ring multi-beam electron-positron collider with a circumference of 237.5 m, beam energy ranging from 1.0 to 2.45 GeV, and luminosity of $~1×10^{33}\ cm^{-2}s^{-1}$. BEPCII has 93 bunches arranged at 8 ns or 2.4 m intervals in each ring. The electron and positron beams collide at a horizontal crossing angle of $\pm$11 mrad at the intersection. 


The Beijing Spectrometer III~(BESIII)~\cite{BESIII:2009fln, Asner:2008nq, BESIII:2020nme}, a spectrometer running at BEPCII, can accurately record the information of final state particles from $e^{+}e^{-}$ collisions. The information is recorded using a two-level trigger system~\cite{Liu:2007mvv}. Level-1~(L1) is hardware-based and level-3~(L3) is software-based, known as the event filter, which operates within the online computer farm. The L1 trigger aims to identify high-quality physics events efficiently and reduce cosmic ray and beam-related background noise. Data stored in the front-end buffers are retrieved once an L1 accept signal is received, approximately 6.4~ms after a collision.
Subsequently, the data is sent to the online computer farm for event processing. The event stream undergoes filtering to further suppress background events through the L3 trigger before saving to the permanent storage device in raw file format.


The BESIII Offline Software System (BOSS)~\cite{Zou:2024pmc, Gaudi}, which is developed with C++ language and object-oriented techniques, uses event data service of Gaudi as data manager to read raw data information. The reconstruction algorithms~\cite{Liang:2009zz, Jia:2010zz, Zhang:2019voa} can retrieve the information from raw data and reconstruct physics information, which will be stored for the subsequent physics analysis. 

Open datasets of experiments, especially from large scientific facilities, play an important role in developing new techniques, physics education, public outreach, and science advocacy~\cite{Guerrieri:2025fhk, Elangovan:2025qhg, Govorkova:2021hqu}. For example, the experiments at the Large Hadron Collider~(LHC) have released some data for unsupervised new physics detection~\cite{Govorkova:2021hqu}. Open datasets are made available in a format that is easy to observe and access directly, allowing users to selectively use specific data to achieve their goals, such as extracting the momentum and initial position of charged particle tracks. 


In this manuscript, we prepare datasets based on BESIII experiment. The main purpose is to provide a small fraction of massive data events collected by BESIII. The dataset can be used to understand the physics in $e^{+}e^{-}$ collisions, and develop event visualization techniques by displaying physics events collected by the BESIII detector. Visualization can help physicists deepen their understanding of some physics processes and has been applied in optimizing the corresponding physics analyses~\cite{Huang:2022wuo, Li:2024pox, You:2017zfr, Zhu:2018mzu, Liao:2024cri, Bianchi:2019can, Tadel:2020hlt, Yuan:2024sns}. In addition, the dataset can effectively support physicists in public outreach and scientific advocacy.
The dataset contains some typical physics processes in $e^{+}e^{-}$ collisions at the~$\tau$-charm energy region, with about 100 events for every process. The data is stored in text format, including the detector hit responses and the reconstructed particle information in each event, making it easy for direct access by the public.


This paper is structured as follows: Sec.~\ref{sec.II} introduces the BESIII detector and data conversion for the open dataset. Sec.~\ref{sec.III}~describe the physics processes and related information in the dataset. Sec.~\ref{sec.IV} presents the technical validation, usage notes, and code availability.

\section{EXPERIMENTAL DESIGN, MATERIALS and Methods} \label{sec.II}

\begin{figure}[!ht]
\centering
\includegraphics[width=1.00\hsize]{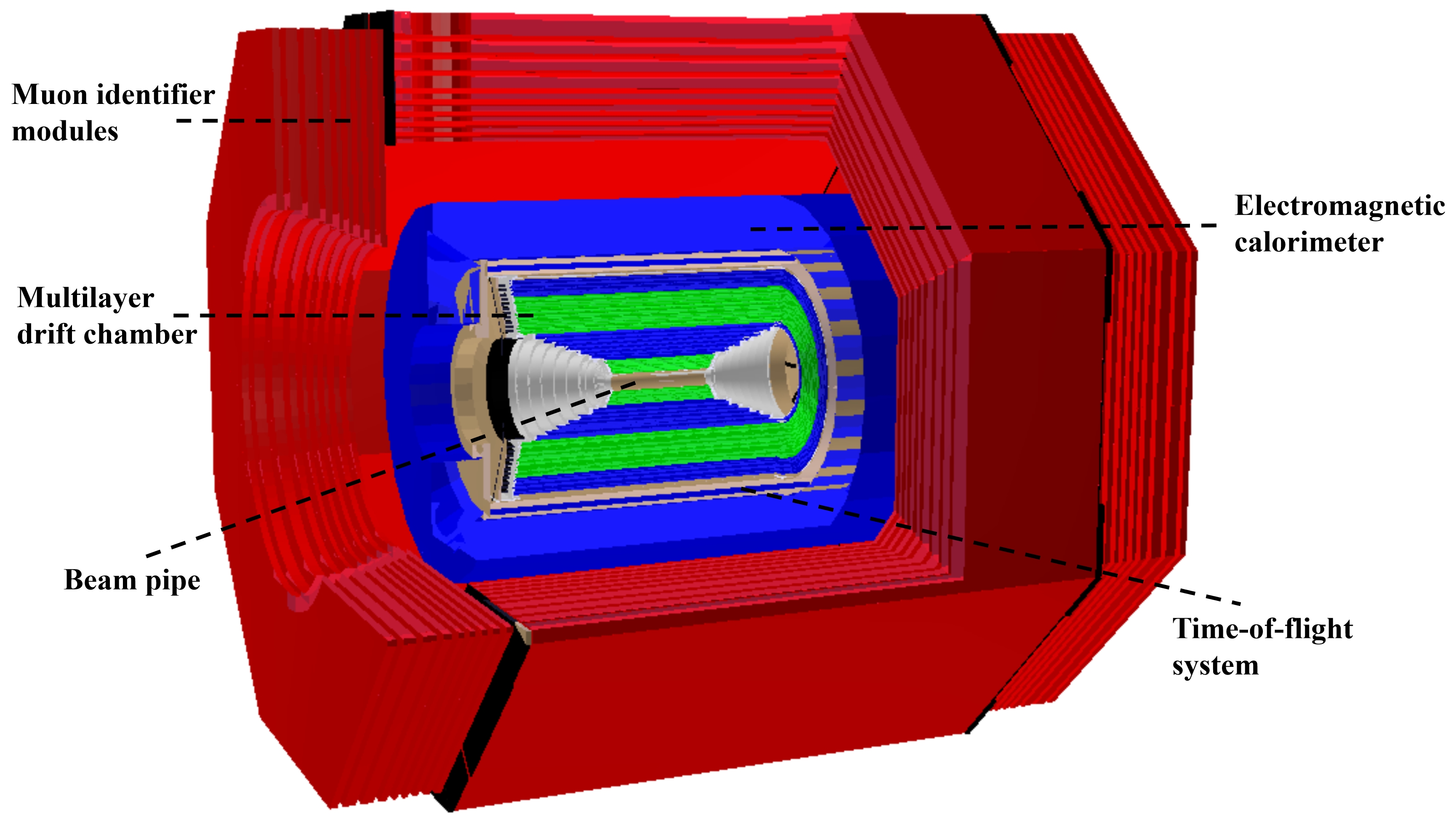}
\caption{The BESIII detector.}
\label{fig:BES_detector}
\end{figure}

The structure of BESIII detector is described below. The cylindrical core consists of a helium-based multilayer drift chamber (MDC), a plastic scintillator time-of-flight system (TOF), and a CsI(Tl) electromagnetic calorimeter (EMC), which are all enclosed in a superconducting solenoidal magnet (SSM) providing a 1.0~T magnetic field. The solenoid is supported by an octagonal flux-return yoke with resistive plate counter muon identifier modules (MUC) interleaved with steel. The acceptance of charged particles and photons is 93\% over $4\pi$ solid angle. The charged particle momentum resolution at 1 GeV/c is 0.5\%, and the $dE/dx$ resolution is 6\% for the electrons from Bhabha scattering. The EMC measures photon energies with a resolution of 2.5\% (5\%) at 1 GeV in the barrel (end-cap) region. The time resolution of the TOF barrel region is 68~ps, while that of the end cap is 110~ps. Visualization of the BESIII detector is shown in Fig.~\ref{fig:BES_detector}. More detailed description of the BESIII detector is available in Ref.~\cite{BESIII:2009fln}.

\begin{figure}[!ht]
\centering
\includegraphics[width=0.70\hsize]{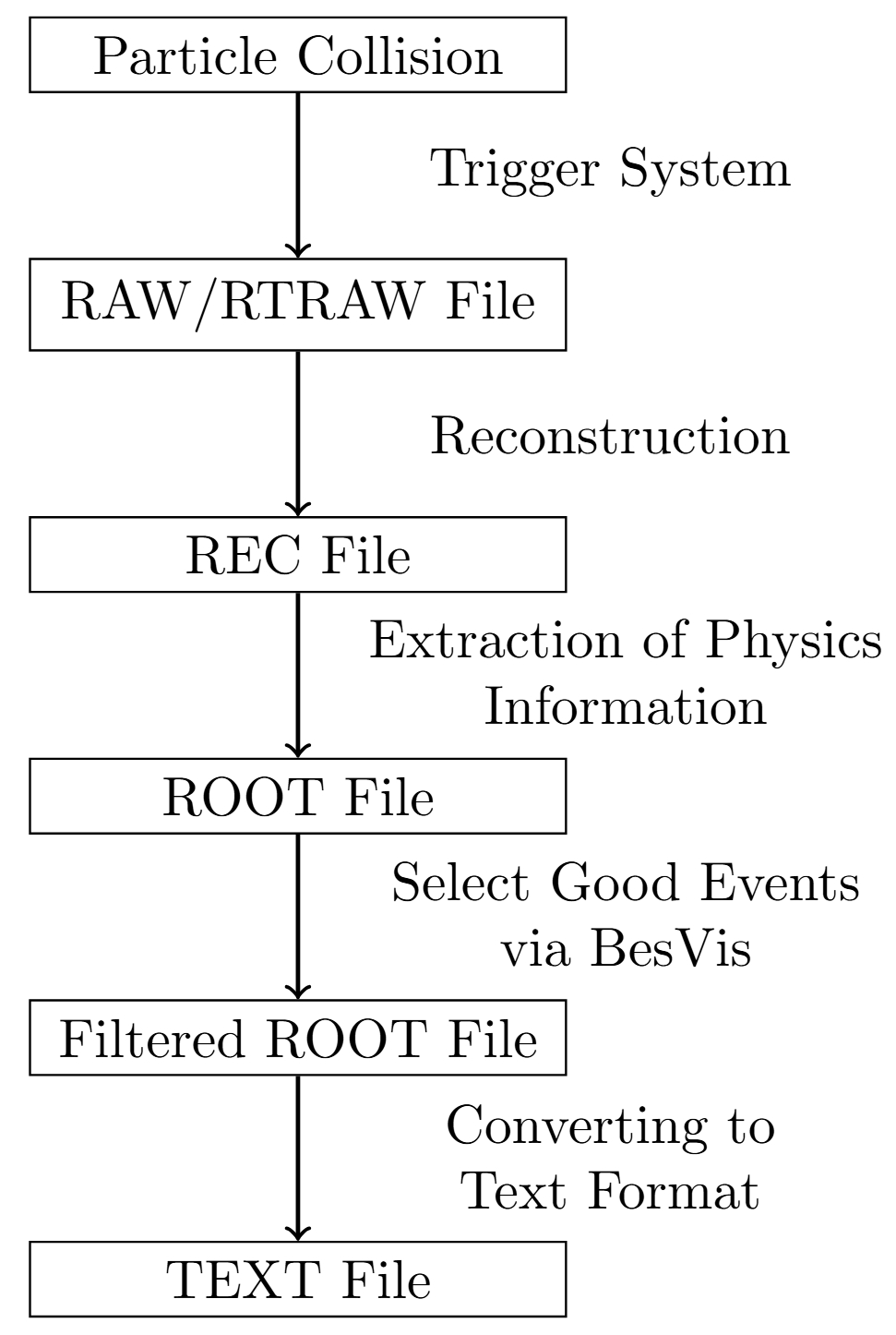}
\caption{The data conversion flow of BESIII event files.}
\label{fig:txtConversion}
\end{figure}

The open dataset is provided in text format, and the data conversion flow is shown in Fig.~\ref{fig:txtConversion}. First, it originates from the raw files of real $e^{+}e^{-}$ collisions or ROOT~\cite{Brun:1997pa} raw~(RTRAW) files generated from the BESIII simulated physics processes. These events pass through the trigger system and are saved on disk~\cite{Liu:2007mvv}. Second, the raw information can be converted into physics information through reconstruction and saved in reconstruction~(REC) format files containing both raw and reconstructed information. These REC files can be effectively viewed using the visualization software BesVis~\cite{Huang:2018fpe} provided by BESIII. 
Third, the event information in the REC files is used to extract physics information with the analysis package, such as the momentum of charged particles, energy deposit in the EMC, and particle penetration length in the MUC. This information is saved in ROOT files. Next, some classic and interesting events can be displayed using BesVis with their event and run number recorded, allowing the selection of good events to be saved in a new filtered ROOT file. Finally, the event information in the filtered ROOT file is converted into a text format for the open dataset.

\section{Data Records}\label{sec.III}
The dataset can be accessed via the repository listed in the specifications table.
Table~\ref{tab:dataset_phy_process} lists the physics processes, selected number of events, and data types included in the dataset. The dataset consists of two parts: one is based on Monte Carlo~(MC) simulation samples using the BESIII offline software, and the other is real data collected by the BESIII detector. The events in the dataset are saved in text format with hit and track information from the detector response. The contents of the dataset will be introduced in the following parts.

\begin{table}[!ht]
\caption{\label{tab:dataset_phy_process}The physics processes, number of events, and data types in the dataset.}
\begin{tabular*}{\hsize} {@{\extracolsep{\fill} } llll}
\toprule
S.N. & Physics processes & Events & Types \\
\midrule
1 & $J/\psi\rightarrow e^{+}e^{-}$     & 128 & MC \\
2 & $J/\psi\rightarrow \mu^{+}\mu^{-}$ & 126 & MC \\
3 & $J/\psi\rightarrow \pi^{+}\pi^{-}\pi^{0}$ & 155 & MC \\
4 & $D^{0}\bar{D}^{0},\ D^{0}\rightarrow K^{-}\pi^{+}, \bar{D}^{0}\rightarrow K^{+}\pi^{-}$ & 122 & MC \\
5 & $D^{+}D^{-},\ D^{+}\rightarrow K^{-}\pi^{+}\pi^{+},\ D^{-}\rightarrow K^{+}\pi^{-}\pi^{-}$ & 140 & MC \\
6 & $\Lambda^{+}_{c}\bar{\Lambda}^{-}_{c},\ \Lambda^{+}_{c}\rightarrow pK^{-}\pi^{+},\ \bar{\Lambda}^{-}_{c}\rightarrow \bar{p}K^{+}\pi^{-}$& 113 & MC \\
7 & $\Lambda^{+}_{c}\rightarrow ne^{+}\nu_{e}$ & 104 & MC \\
8 & $e^{+}e^{-}\rightarrow\pi^{+}\pi^{-}J/\psi\ (Z_{c}(3900))$ & 10 & Data \\
9 & $e^{+}e^{-}\rightarrow e^{+}e^{-}$     & 141 & Data \\
10 & $e^{+}e^{-}\rightarrow \mu^{+}\mu^{-}$     & 121 & Data \\
11 & $e^{+}e^{-}\rightarrow \gamma\gamma$     & 150 & Data \\
12 & 3097 MeV     & 150 & Data \\
13 & 3686 MeV     & 150 & Data \\
14 & 3773 MeV     & 144 & Data \\
15 & Cosmic ray     & 10 & Data \\
\bottomrule
\end{tabular*}
\end{table}

Since some classical physics processes are submerged in the huge background events in real data, it is not easy to select them directly. However, the behavior of MC events is close to real data from event display and can help us determine whether the rare physics events are signals or not. The MC samples include three major decay modes with large branching ratios for the ground state of vector charmonium $J/\psi$, such as the electromagnetic decay $J/\psi\rightarrow e^{+}e^{-}/\mu^{+}\mu^{-}$, and the hadronic decay $J/\psi\rightarrow\pi^{+}\pi^{-}\pi^{0}$~\cite{ParticleDataGroup:2024cfk}, in which $J/\psi\rightarrow e^{+}e^{-}$ is the decay channel where the charm quark was discovered in 1974~\cite{E598:1974sol, Bacci:1974za}. Additionally, the decay modes of charmed meson $D$, such as $D^{0}\rightarrow K^{-}\pi^{+}$, $\bar{D}^{0}\rightarrow K^{+}\pi^{-}$, $D^{+}\rightarrow K^{-}\pi^{+}\pi^{+}$, and $D^{-}\rightarrow K^{+}\pi^{-}\pi^{-}$~\cite{BESIII:2018apz, CLEO:2013rjc}, are provided as they are the main single-tag reconstruction channels in $D$ meson studies widely used in charm physics analysis. Correspondingly, the main decay channel of the lightest charmed baryon $\Lambda^{+}_{c}$, $\Lambda^{+}_{c}\rightarrow pK^{-}\pi^{+}$~\cite{BESIII:2015bjk}, is provided. The dataset also contains the semi-leptonic process $\Lambda^{+}_{c}\rightarrow ne^{+}\nu_{e}$, and the $\bar{\Lambda}^{-}_{c}$ is reconstructed by some main decay channels~\cite{BESIII:2024mgg}. The neutron in the final states has a clear signature in the EMC with a distinct hit distribution. This physics process is also applied in the neutron identification study at BESIII based on deep learning techniques~\cite{BESIII:2024mgg}.

The BESIII detector has collected a large amount of real data at the $\tau$-charm energy region, with the highest integrated luminosity at the energy points of 3097~MeV, 3686~MeV, and 3773~MeV, which correspond to the production thresholds of $J/\psi$, $\psi(3686)$, and $D\bar{D}$, respectively.
The dataset includes real data at these energy points, with approximately 150 events at each energy point, as well as cosmic rays backgrounds and Quantum Electrodynamics~(QED) processes, such as $e^{+}e^{-}\rightarrow e^{+}e^{-}/\mu^{+}\mu^{-}/\gamma\gamma$. 
Additionally, a flagship discovery from real data, the four-quark state $Z_{c}(3900)$ decay events~\cite{BESIII:2013ris, Belle:2013yex}, is also provided in the dataset. 
For some events in the detector, the trajectories of particles appear to resemble the shape of $\psi$, which are widely used in outreach and journal cover images~\cite{Huang:2021xte}.

The filenames in the dataset follow a style that includes the data type, physics process, and recorded information, such as ``Data\_Zc3900\_track\_hit.txt''. Each text file contains detailed reconstructed track and hit information for every event, type setting sequentially as follows: run number, event number, MDC track, MDC hit, TOF track, TOF hit, EMC track, EMC hit, MUC track, and MUC hit. Track and hit contain a variety of digital information, such as momentum, energy, and detector identifier, etc. The definitions of digits in text files are stored in the specification document in the dataset named ``Parameters\_definition.txt''. Users can refer to the document for detailed data information.

\section{Recommended repositories to store and find data}\label{sec.IV}

With the approval from the BESIII collaboration, the dataset has been made public via Science Data Bank with the link: https://doi.org/10.57760/sciencedb.21486. Users can get more information from the BESIII official website (http://bes3.ihep.ac.cn/).

\section*{Technical Validation}

The dataset records a few important physics processes as well as major background events. They have some remarkable features from visualization. Herein, we present various visualization techniques to demonstrate the validity of the dataset.

\begin{figure}[!ht]
\centering
\includegraphics[width=0.95\hsize]{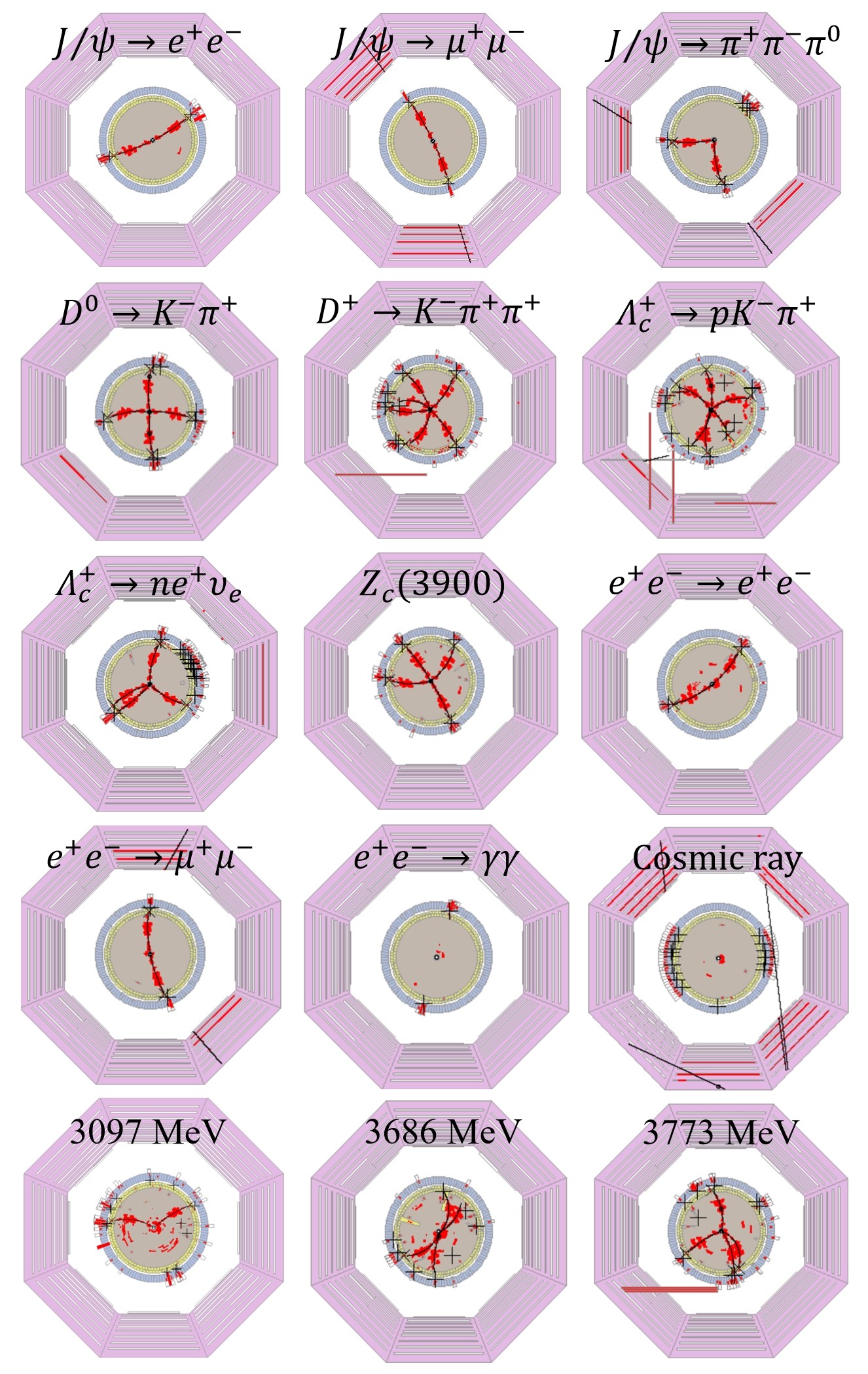}
\caption{Various physics processes displayed with BesVis in X-Y view. From the inside out, the sub-detectors are MDC, TOF, EMC, and MUC.}
\label{fig:BesVis_2D}
\end{figure}

The physic processes in the dataset can be visualized using BesVis, as shown in Fig.~\ref{fig:BesVis_2D}. It has an X-Y cross section view of the BESIII detector, where the tracks and hits can be intuitively displayed. 

\begin{figure}[!ht]
\centering
\includegraphics[width=0.95\hsize]{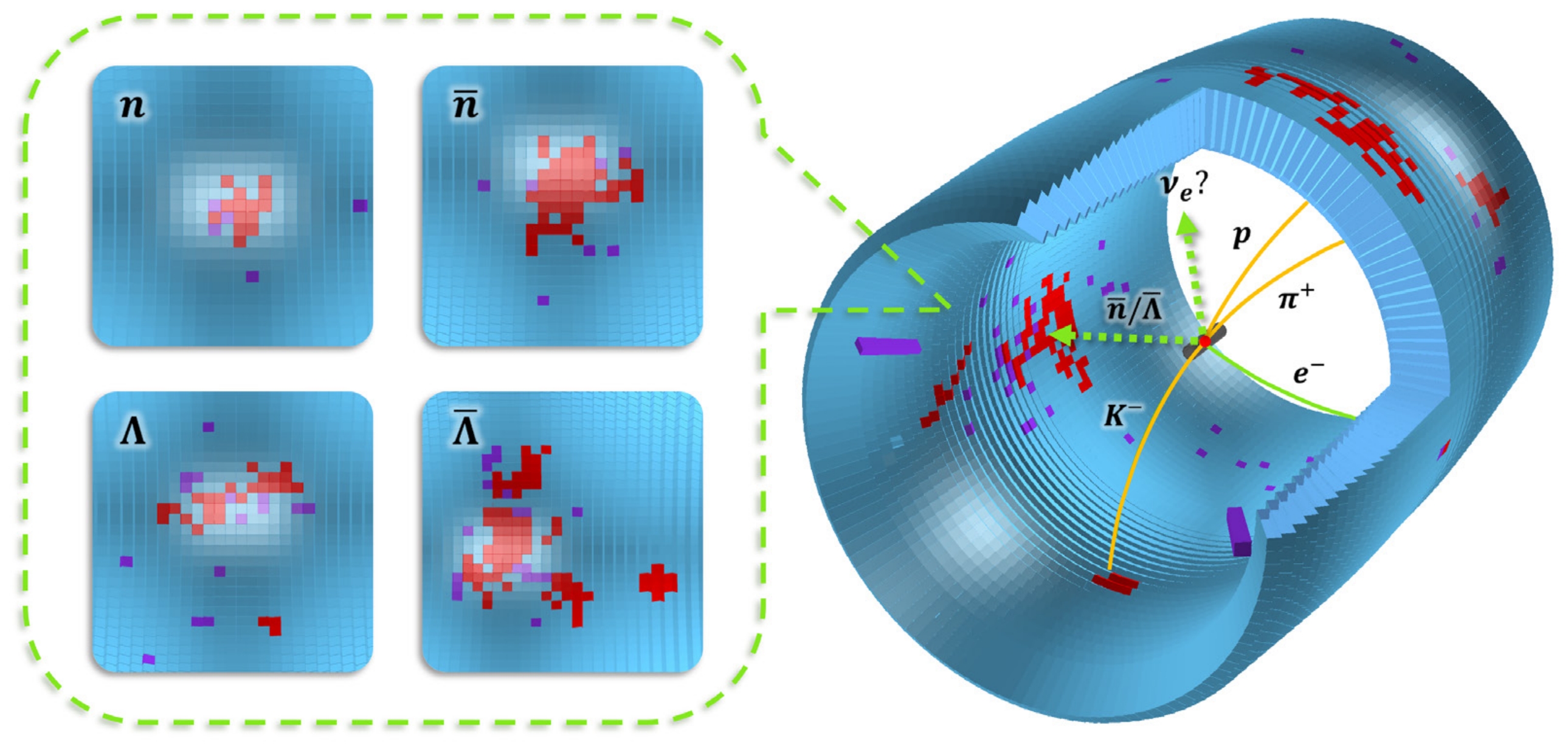}
\caption{Visualization of a $\Lambda^{+}_{c}\rightarrow ne^{+}\nu_{e}$ event using OpenGL~\cite{BESIII:2024mgg}. The blue blocks represent the EMC crystals, and the red blocks indicate the hits of particles with energy deposited in the EMC.}
\label{fig:BesVis_3D}
\end{figure}

In addition to the two-dimensional~(2D) display, a three-dimensional display is also provided using OpenGL, as shown in Fig.~\ref{fig:BesVis_3D}. It demonstrates a $\Lambda^{+}_{c}\rightarrow ne^{+}\nu_{e}$ event, which contains four charged tracks and a clear cluster of EMC hits from neutron energy deposit.

\begin{figure}[!ht]
\centering
\includegraphics[width=0.95\hsize]{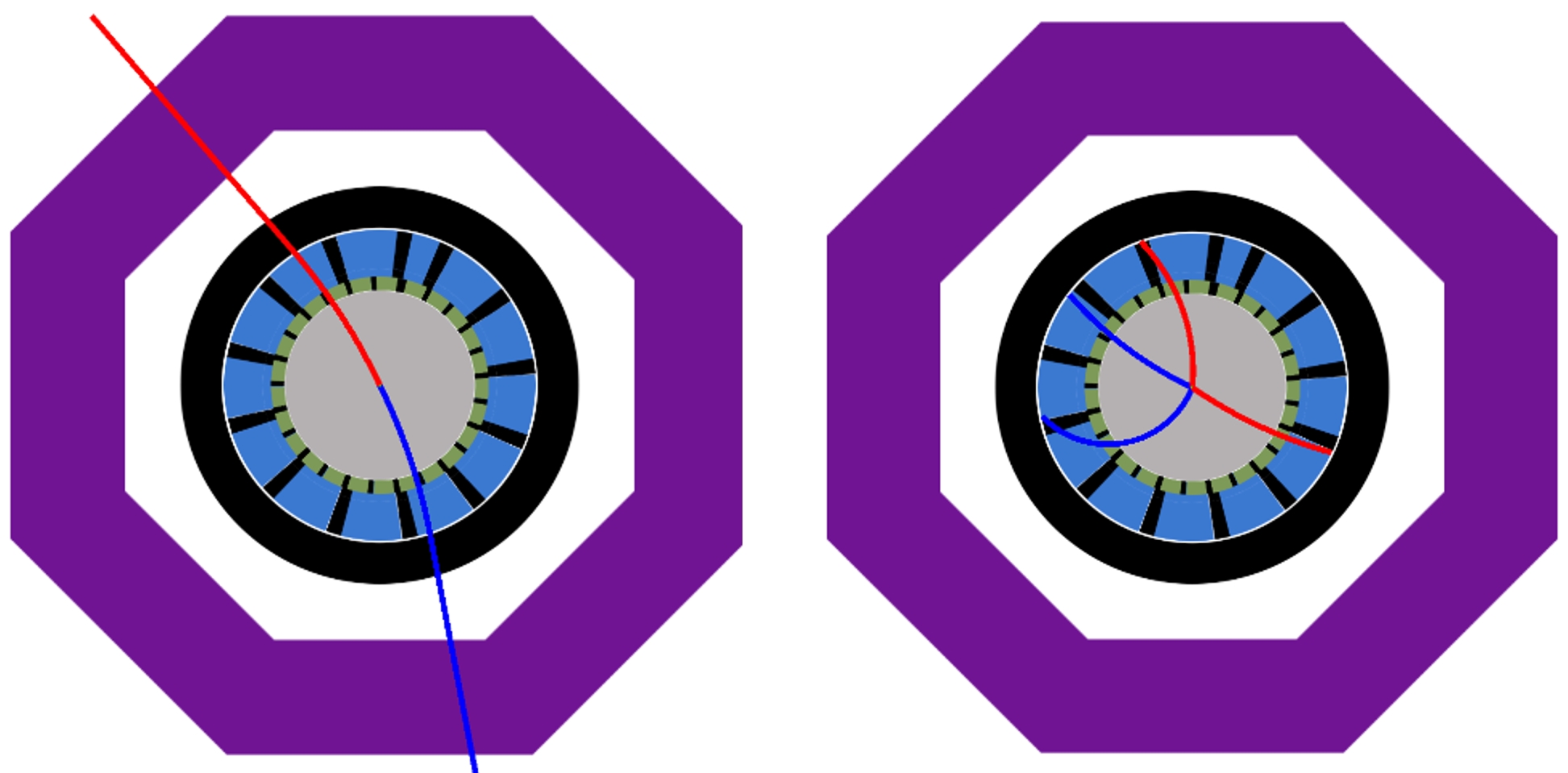}
\caption{Unity-based visualization of a $J/\psi\rightarrow\mu^{+}\mu^{-}$ event and a $Z_{c}(3900)$ event. The red and blue curves represent positively and negatively charged particle tracks, respectively. From the inside out, the subdetectors are MDC, TOF, EMC, SSM, and MUC.}
\label{fig:Unity_2D}
\end{figure}

Moreover, some new techniques of visualization have been developed to display these physics processes. For example, Unity-based event display has been used in BESIII~\cite{Huang:2022wuo, Li:2025ybp}, JUNO~\cite{Zhu:2018mzu}, and other experiments~\cite{Yuan:2024sns}. It has the features of local running, multi-platform deployment, and better visualization effects. Fig.~\ref{fig:Unity_2D} shows the 2D display of a $J/\psi\rightarrow\mu^{+}\mu^{-}$ event, where the muons penetrate the MUC, and a $Z_{c}(3900)$ event, which shows the approximate $\psi$-shaped tracks.

\section*{Usage Notes}

We propose three potential applications with this dataset.

1)~The real data contains not only $J/\psi$, $\psi(3686)$ and $D\bar{D}$ production events, but also continuum backgrounds including QED processes, cosmic ray backgrounds, and other events. The dataset can be used to study the characteristics of different processes.

2)~The dataset has full detector response information at the hit level and track level. Both MC simulation data and real data are provided.
It can be used to study the detector response in simulation and track reconstruction algorithms for development of data processing techniques in offline software.

3)~Visualization plays a crucial role in every aspect of high energy physics experiments. The dataset describes full event information in $e^{+}e^{-}$ collisions. Combined with the BESIII detector geometry, it can be used to develop event display software with various new visualization techniques, not only for physics researches, but also for physics education and public outreach.




\section*{Code availability}

The dataset is publicly available and can be accessed without restrictions. The authors encourage the use of the dataset for visualization research, education, and public outreach. The code is hosted on a public repository, ensuring transparency and facilitating community contributions.



\section*{Author contributions statement}

S.S. Sun and Z.Y. You conceived the experiment, M.H. Liao and X.N. Wang conducted the experiment, M.H. Liao and J.S. Liu analysed the results. All authors reviewed the manuscript. 

\section*{Competing interests} 

The authors declare that they have no conflict of interest.

\end{document}